\documentclass[10pt,showkeys,showpacs,aps,preprintnumbers]{revtex4}

\usepackage{graphicx}
\usepackage{amssymb}
\usepackage{bm}
\usepackage[sumlimits,intlimits]{amsmath}
\usepackage{amsfonts,amssymb}
\usepackage{amsmath}
\usepackage{bm}
\usepackage{graphics}
\usepackage{graphicx}
\usepackage{textcomp}
\usepackage{color}
\usepackage{bm}
\begin{document}
\baselineskip 16pt
\title{Cosmological volume acceleration in dust epoch:
using scaling solutions and variable cosmological term $\Lambda(t)$ within an anisotropic cosmological model.}
\author{J. Socorro$^1$}
\email{socorro@fisica.ugto.mx}
\author{S. P\'erez-Pay\'an$^2$}
\email{saperezp@ipn.mx}
\author{Abraham Espinoza-Garc\'ia$^2$}
\email{aespinoza@ipn.mx}
\author{Luis Rey D\'iaz-Barr\'on$^2$}
\email{lrdiaz@ipn.mx}

 \affiliation{$^1$Departamento de  F\'{\i}sica, DCeI, Universidad
de Guanajuato-Campus Le\'on,
 C.P. 37150, Le\'on, Guanajuato, M\'exico\\
$^2$ Unidad Profesional Interdisciplinaria de Ingenier\'ia Campus Guanajuato del Instituto Polit\'ecnico Nacional.\\
    Av. Mineral de Valenciana \#200, Col. Fraccionamiento Industrial Puerto Interior, C.P. 36275, Silao de la Victoria, Guanajuato,
    M\'exico.
}%
\begin{abstract}
Under the premise that the current observations of the cosmic
microwave background radiation set a very stringent limit to the
anisotropy of the universe, we consider an anistropic model in the
presence of a barotropic perfect fluid and a homogeneous scalar
field, which transits to a flat FRW cosmology for late times in a
dust epoch, presenting an accelerated volume expansion.
Furtheremore, the scalar field is identified with a varying
cosmological term via $V(\phi(t))=2\Lambda(t)$. Exact solutions to
the  EKG system are obtained by proposing an anisotropic extension
of the scaling solutions scenario:
 $\rm\rho\sim \eta^{-n},\ \rho_\phi\sim \eta^{-m}$, with $\rm\eta^3=a_1a_2a_3$ the volume function of
  the anistropic model ($\rm a_1,\, a_2,\, a_3$ being the scale factors).
 \end{abstract}

 \maketitle

%
\section{introduction}
A number of cosmological observations \citep{Knop et al:2003, Riess
et al:2004, Tegmark et al:2004, Spergel et al:2007} indicate
altogether that our universe is presently under a phase of
accelerated expansion. Such stage is characterized by  which is
popularly known as dark energy, whose simplest incarnation is the so
called cosmological constant. Furthermore, observations of the
cosmic microwave background radiation (CMB) indicate that the
universe feature a high degree of homogeneity and isotropy. Of
course, any theoretical model must at least be in accord with these
observations as a first step regarding its viability.

Models with different decay laws for the variation of the
cosmological term have been investigated in a non-covariant way in
\citep{Chen:1990, Abdel:1990,Pavon:1991, Carvalho et al:1992,
Kalligas et al:1992, Maia:1994, Lima:1994, Trodden:1996, Abdel:1994,
Birkel:1997, Silveira:1997, Starobinsky:1998, Overduin:1998,
Vishwakarma:2000, Vishwakarma:2001, Arbab2001, Arbab2003, Arbab2004,
Cunha:2004, Carneiro:2005, Fomin:2005, Stefancic:2005, Sola:2006,
Pradhana2007, Jamil:2011, Mukhopadhyay:2011}. In \citep{Fomin:2005}
several evolution relations for $\Lambda$ (which many authors have
used) can be found. Also, in Ref. \citep{Overduin:1998}, a table
with such relations and the corresponding references where they were
considered is presented. There are other alternatives, e.g.,
\citep{Ray et al:2011,Shahalam et al:2015} where the authors explore
a realization of time-dependent $\Lambda$ as a positive pressure.

The current observations of the CMB set a very stringent limit to
the anisotropy of the universe \citep{Martinez:1995}, therefore, if
anistropic cosomological models are to be considered, it is
important to deal with those which isotropize during evolution
\citep{Belinskii:1972, Folomeev:2000}. Hence, there is a natural
desire to build an anisotropic cosmological model having some of the
advantages which the flat Friedmann model present, and to analyze
the possibility of its isotropization. One of the most popular
conditions to establish isotropization is that of asymptotic
constancy of the parameters which are related to the anisotropy of
the model. In the present work, a Misner's parametrization variant
enables us to explicitly separate the anisotropic part (shape) from
the isotropic one (volume), and we show that such isotropization
condition is met for the dust epoch. Some authors \citep{Bylan:1998,
Bali:2002, Bali:2011} have considered anisotropy in studies of
primordial inflation, and in order to find exact solutions to the
Einstein-Klein-Gordon (EKG) equations it was necessary for the
potential to approach asymptotically to a fixed value. The scalar
field that we obtain in the present investigation also exhibits this
behavior, nonetheless, to our knowledge, it has not been studied in
the literature. In addition, anisotropic cosmological models have
been treated in this formalism from different points of view.
\citep{Aroonkumar:1993, Aroonkumar:1994, Arbab:1997, Singh:1998,
Kumar:2001, Pradhan:2003, Pradhan:2007, Pradhan:2009, Pandey:2003,
PradhanPandey:2006, Saha:2006, Pradhan2007, Pradhan2008,
Pradhan2012, Pradhan2015, Carneiro:2005, Esposito et al:2007,
Kumar:2007, Bal:2008, Belinchon:2008, Singh et al:2008, Singh et
al:2013, Shen:2013, Tripathy:2013, Ansary:2013}.

Recently the Bianchi type I anisotropic cosmological model has been
used to clarify the recent observations that indicate very tiny
variations in the intensity of microwaves coming from different
directions in the sky, employing the energy density approach mixed
with some number of data in H(z) and the distance modulus of
supernovae to constraint the cosmological parameters
\citep{Amirhashchi:2018, Amirhashchi:2019, Akarsu et al:2019,
    Goswami et al:2020}. Also, other type of anisotropic cosmological
models are employed \citep{Zia et al:2019}. However, these authors
do not indicate which scalar potential is used in order to explain
the accelerated expansion of the universe, nor the isotropization of
this anisotropic cosmological models, nor the value acquired by the
barotropic parameter $\omega_{DE}$ in the phantom region. In
particular, in reference \citep{Amirhashchi:2018}, the author states
that, possibly, the average anisotropic parameter $\rm \overline
A_m$ could be of the form $\rm \overline A_m=-6K$, with $\rm K$ a
negative constant. In our work, we found that the average
anisotropic parameter have an upper limit, being $\rm \overline A_m
\leq 0.54$, which is consistent with the above remark.

In the present investigation, we introduce a combination of results
by making two extensions. The first one has to do with performing an
anisotropic generalization of a particular scaling solutions
scenario \citep{Joyce:1998, Liddle:1998, Copeland et al:1998} known
as the ``attractor solution". In the usual (isotropic) scaling
solutions scenario the energy density $\rho$ of the barotropic fluid
and that of the scalar field are related to the FRW scale factor $A$
by $\rm\rho\sim A^{-n}$, $\rm\rho_{\phi}\sim A^{-m}$. The particular
``atractor solution'' corresponding to the choice $\rm m=n$, which
translates to proportionality among the energy densities: $\rm
\rho_\phi=m_\phi \rho$ \citep{Liddle:1998}. This scaling solutions
scheme adapted to the anistropic Bianchi type I model enables us to
find exact solutions for the gravitational potentials and the scalar
field in a dust epoch. The second idea is to extend the usual
identification $\rm V(\phi)=2\Lambda$ ($\Lambda$ being the
cosmological \textit{constant}) to the dynamical case, i.e., $\rm
V(\phi(t))=2\Lambda(t)$. Employing the exact solution found for the
scalar field, we find that the cosmological term $\rm \Lambda(t)$ is
a decreasing function of time and that it approaches a small
positive value at the present epoch, which is consistent with
results of recent supernovae Ia observations \citep{Tegmark et
al:2004}.

As a consequence of scaling solutions, the dark content of the
universe can be accounted for by a dimensionless parameter $\rm
m_\phi$, where (considering the approximate percentages $5\%$ for
ordinary matter, $25\%$ for dark matter and $70\%$ for dark energy)
$\rm m_\phi=5$ for dark matter, and $\rm m_\phi=14$ for dark energy.
The evolution of the isotropic volume $\eta^3$ is anticipated to be
different for each of these values, a more significant growth for
$\rm m_\phi=14$ than for $\rm m_\phi=5$ is expected. However, the
nature of the composition of the scalar field is unknown since this
type of matter cannot be detected by the usual methods. Recently,
one of the authors presented an analysis in a covariant way
\citep{Socorro et al:2015}, where the cosmological term of the FRW
model is used, obtaining the corresponding scalar potential in
different scenarios as well as its relationship with the
time-dependent cosmological term.

In the present treatment we take into  account the corresponding
Lagrangian density with a scalar field
\begin{equation}
 \rm {\cal L}[g,\phi]=\rm \sqrt{-g}\left(R-\frac{1}{2}g^{\mu \nu}\nabla_\mu\phi\nabla_\nu\phi
 +V(\phi)\right)+\sqrt{-g} {\cal L}_{matter}\label{lagra}
\end{equation}
where $\rm R$ is the Ricci scalar, ${\cal L}_{matter}$ corresponds
to a barotropic perfect fluid, $\rm p=\omega \rho$, $\rho$ is the
energy density, $\rm p$ is the pressure of the fluid in the
co-moving frame and $\omega$ is the barotropic constant. The
corresponding variation of (\ref{lagra}), with respect to the metric
and the scalar field, results in the EKG field equations
\begin{eqnarray}
\rm G_{\alpha\beta}&=&\rm
-\frac{1}{2}\left(\nabla_\alpha\phi\nabla_\beta\phi-\frac{1}{2}g_{\alpha\beta}
g^{\mu\nu}\nabla_\mu\phi\nabla_\nu\phi\right)+\frac{1}{2}g_{\alpha\beta}V(\phi)-8\pi GT_{\alpha\beta}, \label{camrel}\\
\rm \square\phi-\frac{\partial V}{\partial\phi}&=&0.\label{klein}
\end{eqnarray}
From  (\ref{camrel}) it  can be deduced  that the energy-momentum
tensor associated with the scalar field is
\begin{equation}
\rm 8 \pi G T^{(\phi)}_{\alpha\beta}=\rm
\frac{1}{2}\left(\nabla_\alpha\phi\nabla_\beta\phi-\frac{1}{2}g_{\alpha\beta}
g^{\mu\nu}\nabla_\mu \phi\nabla_\nu\phi\right)
-\frac{1}{2}g_{\alpha\beta}V(\phi),
\end{equation}
and the corresponding tensor for a barotropic perfect fluid  becomes
\begin{equation}
\rm T_{\alpha \beta}= \left(p+\rho\right) u_\alpha u_\beta +
g_{\alpha \beta} p,
\end{equation}
here $\rm u_\alpha$ is the four-velocity in the co-moving  frame,
and the barotropic equation of state is $\rm p=\gamma \rho$.

\subsection{ Misner's parametrization variant}
The line element for the anisotropic cosmological Bianchi type I
model in the Misner's parametrization
\begin{eqnarray}
\rm ds^2 &=& \rm -N^2 dt^2 +a_1^2 dx^2 + a_2^2 dy^2 + a_3^2 dz^2, \nonumber \\
 &=&\rm -N^2 dt^2 + e^{2\Omega}\left[e^{2\beta_++2\sqrt{3}\beta_-}dx^2 +e^{2\beta_+-2\sqrt{3}\beta_-} dy^2 + e^{-4\beta_+} dz^2\right],\label{biachi_I_misner}
 \end{eqnarray}
 where $\rm a_i$ ($\rm i=1,2,3$) are the scale factor on directions $\rm (x,y,z)$, respectively, and N is the lapse function.
 For convenience, and in order to carry out the analytical calculations, we consider the following representation for the line
 element (\ref{biachi_I_misner})
\begin{equation}
\rm ds^2 = -N^2 dt^2 + \eta^2\left[m_1^2 dx^2 +m_2^2 dy^2 + m_3^2
dz^2\right], \label{bianchi}
 \end{equation}
 where the relations between both representations are given by
 \begin{eqnarray}
 \rm \eta&=& \rm e^{\Omega}, \nonumber\\
 \rm m_1 &=& \rm e^{\beta_+ + \sqrt{3} \beta_-}, \qquad \frac{\dot m_1}{m_1}=\dot \beta_+ + \sqrt{3} \dot \beta_-, \nonumber\\
 \rm m_2&=& \rm e^{\beta_+ - \sqrt{3} \beta_-}, \qquad \frac{\dot m_2}{m_2}=\dot \beta_+ - \sqrt{3} \dot \beta_-,\nonumber\\
 \rm m_3&=& \rm e^{-2\beta_+ }, \qquad \qquad \frac{\dot m_3}{m_3}=-2\dot \beta_+,\nonumber
\end{eqnarray}
 where $\eta$ is a function that has information regarding the isotropic scenario and the $\rm m_i$ are dimensionless functions
 that has  information about the anisotropic behavior of the universe, such that
\begin{eqnarray}
\rm \rm \Pi_{i=1}^{3}m_i&=&1, \nonumber\\
\rm  \Pi_{i=1}^{3}a_i&=&\rm \eta^3, \nonumber\\
\rm  \sum_{i=1}^3 \frac{\dot m_i}{m_i}&=&\rm 0, \label{mi}
\end{eqnarray}
act as constraint equations for the model.

\section{Field equations of the Bianchi type I cosmological model}

In this section we present the solutions of the field equations for
the anisotropic cosmological model, considering the temporal
evolution of the scale factors with barotropic fluid and standard
matter. The solutions obtained already consider the particular
choice of the Misner-like transformation discussed lines above.
Using the metric (\ref{bianchi}) and a co-moving fluid, equations
(\ref{camrel}) take the following form
\begin{eqnarray}
\rm \left( \begin{tabular}{c} 0\\0 \end{tabular} \right)&& \rm
 \frac{\dot m_1}{Nm_1} \frac{\dot m_2}{Nm_2}+\frac{\dot m_2}{Nm_2}\frac{\dot m_3}{Nm_3}+\frac{\dot m_1}{Nm_1}\frac{\dot m_3}{Nm_3}
  +3 \left(\frac{\dot \eta}{N\eta}\right)^2-8\pi G\rho
 -\frac{1}{2}\left(  \frac{1}{2} \frac{\dot \phi^2}{N^2}+V(\phi)\right)=0,\label{0,0}\\
\rm \left( \begin{tabular}{c} 1\\1 \end{tabular} \right)&&\rm
-\frac{\dot N}{N^2}\left[\frac{\dot m_2}{Nm_2}+\frac{\dot m_3}{Nm_3}+2\frac{\dot \eta}{N\eta}\right]+\frac{\ddot m_2}{N^2m_2}+ \frac{\ddot m_3}{N^2m_3}+\frac{\dot m_2}{Nm_2}\frac{\dot m_3}{Nm_3}+2\frac{\ddot \eta}{N^2\eta}+\left(\frac{\dot \eta}{N\eta}\right)^2+3\frac{\dot \eta}{N\eta}\left[\frac{\dot m_2}{Nm_2}+\frac{\dot m_3}{Nm_3} \right] \nonumber\\
&&\mbox{} \rm +\frac{1}{2}\left(  \frac{1}{2} \frac{\dot
\phi^2}{N^2}-V(\phi)\right)+ 8\pi G P=0,
 \label{1,1}\\
\rm \left( \begin{tabular}{c} 2\\2 \end{tabular}\right)&&\rm
 -\frac{\dot N}{N^2}\left[\frac{\dot m_1}{Nm_1}+\frac{\dot m_3}{Nm_3}+2\frac{\dot \eta}{N\eta}\right]+\frac{\ddot m_1}{N^2m_1} + \frac{\ddot m_3}{N^2m_3}+\frac{\dot m_1}{Nm_1}\frac{\dot m_3}{Nm_3}+2\frac{\ddot \eta}{N^2\eta}+\left(\frac{\dot \eta}{N\eta}\right)^2 +3\frac{\dot \eta}{N\eta}\left[\frac{\dot m_1}{Nm_1}+\frac{\dot m_3}{Nm_3} \right] \nonumber\\
&&\mbox{} \rm   +\frac{1}{2}\left(  \frac{1}{2} \frac{\dot \phi^2}{N^2}-V(\phi)\right)  + 8\pi G P=0,\label{2,2}\\
\rm \left( \begin{tabular}{c} 3\\3 \end{tabular}\right)&&\rm
  -\frac{\dot N}{N^2}\left[\frac{\dot m_1}{Nm_1}+\frac{\dot m_2}{Nm_3}+2\frac{\dot \eta}{N\eta}\right]+\frac{\ddot m_1}{N^2m_1}+ \frac{\ddot m_2}{N^2m_2}+\frac{\dot m_1}{Nm_1}\frac{\dot m_2}{Nm_2}  +2\frac{\ddot \eta}{N^2\eta}+\left(\frac{\dot \eta}{N\eta}\right)^2 +3\frac{\dot \eta}{N\eta}\left[\frac{\dot m_1}{Nm_1}+\frac{\dot m_2}{Nm_2} \right] \nonumber\\
&&\mbox{} \rm
 +\frac{1}{2}\left(  \frac{1}{2} \frac{\dot \phi^2}{N^2}-V(\phi)\right)  + 8\pi G P=0,\label{3,3}
\end{eqnarray}
here a dot (~$\dot{}$~) represents a time derivative. The
corresponding Klein-Gordon (KG) equation is given by
\begin{equation}
\rm \frac{\dot N}{N}\frac{\dot \phi^2}{N^2}- \frac{\dot \phi \ddot
\phi}{N^2}-3\frac{\dot \eta}{\eta} \frac{\dot \phi^2}{N^2}- \dot
V=0, \label{klein-gordon}
\end{equation}
and the conservation law for the energy-momentum tensor reads
\begin{equation}
\rm 3(\gamma+1)\frac{\dot \eta}{\eta} + \frac{\dot
\rho}{\rho}=0,\label{ordinary}
\end{equation}
with solution $\rm \rho=\rho_\gamma \eta^{-3(1+\gamma)}$ and $\rm
\rho_\gamma$ is an integration constant that depends on the scenario
that is being considered. By setting $\rm 16\pi G \rho_\phi=
\frac{1}{2} \frac{\dot \phi^2}{N^2}-V(\phi)$ and $\rm 16\pi G
P_\phi= \frac{1}{2} \frac{\dot \phi^2}{N^2}+V(\phi)$ as the energy
density and pressure of the scalar field, respectively, and a
barotropic equation of state for the scalar field of the form $\rm
p_\phi=\omega_\phi \rho_\phi$, we can obtain the kinetic energy $\rm
K= \frac{1+\omega_\phi}{1-\omega_\phi} V(\phi)$, therefore, equation
(\ref{klein-gordon}) can be written as
\begin{equation}
\rm \frac{d}{d\tau} \left[ Ln\left(\eta^6
V^{\frac{2}{1+\omega_\phi}} \right) \right]=0,  \rightarrow \quad
V=c_\omega \eta^{-3(1+\omega_\phi)}, \label{potential}
\end{equation}
hence, the scalar field has solutions in quadrature form
\begin{equation}
\rm \Delta \phi = \alpha_\omega \int
\frac{d\tau}{\eta^{\frac{3(1+\omega_\phi)}{2}}},
\label{scalar-field}
\end{equation}
where $\rm c_\omega$ and $\rm \alpha_\omega$ are appropriate
constants for the scenario considered. The relation $\rm
p_\phi=\omega_\phi \rho_\phi$ implies  that $\rm \rho_\phi=\frac{2
c_\omega}{1-\omega_\phi} \eta^{-3(1+\omega_\phi)}\sim \eta^{-m}$
where $\rm m=3(1+\omega_\phi)$. On the other hand, the solution of
the energy-momentum tensor for the perfect fluid ($\rm \nabla_\nu
T^{\mu \nu}=0$), gives $\rm\rho=\rho_\gamma \eta^{-3(1+\gamma)}\sim
\eta^{-n}$, where $\rm n=3(1+\gamma)$, where in principle, the two
barotropic indexes $\rm\omega_\phi$ and $\gamma$ are different. It
is well documented in the literature that in the FRW cosmological
model, for the case $\rm m = n$ the solutions obtained are ``
attractor solutions '' and correspond to the potential of the scalar
field $\phi$ having an exponential behavior. This case has been
studied by several authors, using different methods, with the
potential put by hand \citep{Lucchin:1985, Halliwell:1985,
Burd:1988, Weetterich:1998, Wand et al:1993, Ferreira:1997, Copeland
et al:1998}, in order to understand the evolution of the universe.
It turns out that for anisotropic cosmological models there is no
available literature regarding this topic.

In order to solve the set of equations (\ref{0,0})-(\ref{ordinary}),
we consider the case $\rm{m=n}$, thus, we have that
$\gamma=\omega_\phi$, and found that the energy density of the
scalar field and the energy density of
 ordinary matter must satisfy
the relation $\rho_{\omega_\phi}=\rm m_\phi \rho$, where $\rm
m_\phi$ is a positive constant that gives the proportionality
between the dark matter-energy and ordinary matter, and as a
consequence we find the corresponding potential of the scalar field
for a wide range of values of the barotropic index $\omega_\phi$,
where the temporal solution for standard matter is found.

From here on, we introduce the dimensionless parameter
$\alpha_\phi=1+\rm m_\phi$, which is equal to unity for in absence
of scalar field. To obtain the time dependent solutions of the scale
factors we write the EKG equations as
\begin{eqnarray}
\rm \left( \begin{tabular}{c} 0\\0 \end{tabular} \right)&&
\rm\frac{\dot m_1}{Nm_1} \frac{\dot m_2}{Nm_2}+\frac{\dot m_2}{Nm_2}\frac{\dot m_3}{Nm_3}+\frac{\dot m_1}{Nm_1}\frac{\dot m_3}{Nm_3}+3 \left(\frac{\dot \eta}{N\eta}\right)^2-8\pi G \alpha_\phi \rho =0, \label{00-alfa}\\
\rm \left( \begin{tabular}{c} 1\\1 \end{tabular} \right)&&
\rm -\frac{\dot N}{N^2}\left[\frac{\dot m_2}{Nm_2}+\frac{\dot m_3}{Nm_3}+2\frac{\dot \eta}{N\eta}\right]+\frac{\ddot m_2}{N^2m_2} + \frac{\ddot m_3}{N^2m_3}+\frac{\dot m_2}{Nm_2}\frac{\dot m_3}{Nm_3} +2\frac{\ddot \eta}{N^2\eta}+\left(\frac{\dot \eta}{N\eta}\right)^2 \nonumber\\
&&\mbox{} \rm +3\frac{\dot \eta}{N\eta}\left[\frac{\dot
m_2}{Nm_2}+\frac{\dot m_3}{Nm_3} \right]
 + 8\pi G \gamma \alpha_\phi P=0,\label{11-alfa}\\
\rm \left( \begin{tabular}{c} 2\\2 \end{tabular}\right)&&\rm
 -\frac{\dot N}{N^2}\left[\frac{\dot m_1}{Nm_1}+\frac{\dot m_3}{Nm_3}+2\frac{\dot \eta}{N\eta}\right]+\frac{\ddot m_1}{N^2m_1}+ \frac{\ddot m_3}{N^2m_3}+\frac{\dot m_1}{Nm_1}\frac{\dot m_3}{Nm_3} +2\frac{\ddot \eta}{N^2\eta}+\left(\frac{\dot \eta}{N\eta}\right)^2\nonumber\\
&&\mbox{} \rm +3\frac{\dot \eta}{N\eta}\left[\frac{\dot
m_1}{Nm_1}+\frac{\dot m_3}{Nm_3} \right]
 + 8\pi G \gamma \alpha_\phi P=0,\label{22-alfa}\\
\rm \left( \begin{tabular}{c} 3\\3 \end{tabular}\right)&&\rm
  -\frac{\dot N}{N^2}\left[\frac{\dot m_1}{Nm_1}+\frac{\dot m_2}{Nm_3}+2\frac{\dot \eta}{N\eta}\right]+\frac{\ddot m_1}{N^2m_1}+ \frac{\ddot m_2}{N^2m_2}+\frac{\dot m_1}{Nm_1}\frac{\dot m_2}{Nm_2} +2\frac{\ddot \eta}{N^2\eta}+\left(\frac{\dot \eta}{N\eta}\right)^2\nonumber\\
&&\mbox{} \rm +3\frac{\dot \eta}{N\eta}\left[\frac{\dot
m_1}{Nm_1}+\frac{\dot m_2}{Nm_2} \right]
 + 8\pi G \gamma  \alpha_\phi P=0,\label{33-alfa}
\end{eqnarray}
and the KG equation for this particular case is the same as before
(equation (\ref{klein-gordon}))
\begin{equation}
\rm \frac{\dot N}{N}\frac{\dot \phi^2}{N^2}- \frac{\dot \phi \ddot
\phi}{N^2}-3\frac{\dot \eta}{\eta} \frac{\dot \phi^2}{N^2}- \dot
V=0.  \label{KG-alfa}
\end{equation}
Now, substracting (\ref{11-alfa}) from the component (\ref{22-alfa})
we obtain
\begin{equation}
\rm \frac{\dot N}{N^3}\left[\frac{\dot m_1}{m_1} -\frac{\dot
m_2}{m_2}\right] -\frac{\dot m_1}{Nm_1}\frac{\dot
m_3}{Nm_3}+\frac{1}{N^2}\left[\frac{\ddot m_2}{m_2}-\frac{\ddot
m_1}{m_1}\right] +\frac{\dot m_2}{Nm_2}\frac{\dot
m_3}{Nm_3}+3\frac{\dot \eta}{N^2\eta}\left[\frac{\dot
m_2}{m_2}-\frac{\dot m_1}{m_1}\right]=0, \label{1-2}
\end{equation}
noticing that
\begin{equation*}
\rm \frac{1}{N}\left[ \frac{\dot m_2}{Nm_2} - \frac{\dot m_1}{Nm_1}
\right]^\bullet =\frac{1}{N^2}\left[\frac{\ddot m_2}{m_2} -
\frac{\ddot m_1}{m_1} \right] -\frac{1}{N^2}\left[ \left(\frac{\dot
m_2}{m_2}\right)^2 -\left(\frac{\dot m_1}{m_1}\right)^2\right] +
\frac{\dot N}{N^3}\left[\frac{\dot m_1}{m_1} -\frac{\dot m_2}{m_2}
\right],
\end{equation*}
equation (\ref{1-2}) can be rearranged and written as (where
$^\bullet$ also denotes a time derivative)
$$
\rm \frac{1}{N}\left[ \frac{\dot m_2}{Nm_2} - \frac{\dot m_1}{Nm_1}
\right]^\bullet +3\frac{\dot \eta}{N\eta}\left[\frac{\dot
m_2}{Nm_2}-\frac{\dot m_1}{Nm_1}\right]=0, \label{12}
$$
finally, defining $\rm R_{21}= \frac{\dot m_2}{Nm_2} - \frac{\dot
m_1}{Nm_1}$, the last equation can be casted as $ \rm \frac{\dot
R_{21}}{R_{21}}+3\frac{\dot \eta}{\eta}=0, $ whose solution is given
by
\begin{equation}
\rm R_{21}=\frac{\ell_{21}}{\eta^3},\label{21}
\end{equation}
where $\rm \ell_{21}$ is an integration constant.

When we perform the same procedure with the other pair of equations,
namely, subtracting (\ref{22-alfa}) from the component
(\ref{33-alfa}) one obtains
\begin{equation}
\rm \frac{\dot N}{N^3}\left[\frac{\dot m_2}{m_2} -\frac{\dot
m_3}{m_3}\right] -\frac{\dot m_2}{Nm_2}\frac{\dot
m_1}{Nm_1}+\frac{1}{N^2}\left[\frac{\ddot m_3}{m_3}-\frac{\ddot
m_2}{m_2}\right]
 +\frac{\dot m_1}{Nm_1}\frac{\dot m_3}{Nm_3}+3\frac{\dot \eta}{N^2\eta}\left[\frac{\dot m_3}{m_3}-\frac{\dot m_2}{m_2}\right]=0, \label{2-3}
\end{equation}
which has the same structure as equation (\ref{1-2}). Proceeding in
the same manner as we did above, we define $\rm R_{32}=\frac{\dot
m_3}{Nm_3} - \frac{\dot m_2}{Nm_2}$, obtaining a differential
equation  whose solution is analogous to (\ref{21}),
\begin{equation}
\rm R_{32}=\frac{\ell_{32}}{\eta^3},\label{32}
\end{equation}
where $\rm \ell_{32}$ is an integration constant. And lastly,
subtracting (\ref{11-alfa}) from (\ref{33-alfa}) we get
\begin{equation}
\rm R_{13}=\frac{\ell_{13}}{\eta^3},\label{13}
\end{equation}
$\rm\ell_{13}$ is also a constant that comes from integration, these
three constants satisfy $\rm \ell_{21}+\ell_{32} +\ell_{13}=0$.

From equation (\ref{21}) we have that
\begin{equation}
\rm 2 \frac{\dot m_2}{Nm_2}-\frac{\dot m_1}{Nm_1}-\frac{\dot
m_2}{Nm_2}=\frac{\ell_{21}}{\eta^3},
\end{equation}
and if we use the constraints from (\ref{mi}), the last equation
reduces to
\begin{equation}
\rm 2 \frac{\dot m_2}{Nm_2}+\frac{\dot
m_3}{Nm_3}=\frac{\ell_{21}}{\eta^3},
\end{equation}
finally, using equation (\ref{32}) as a last step we get
\begin{equation}
\rm 3 \frac{\dot
m_2}{Nm_2}+\frac{\ell_{32}}{\eta^3}=\frac{\ell_{21}}{\eta^3}.
\end{equation}
In order to investigate the solution for the last equation we cast
it in the following form
\begin{equation}
\rm \frac{\dot
m_2}{Nm_2}=\frac{\ell_{21}-\ell_{32}}{3\eta^3}=\frac{\ell_2}{\eta^3},\label{m2}
\end{equation}
where $\ell_2=\frac{\ell_{21}-\ell_{32}}{3}$. The other components
can be obtained in a similar fashion, which read
\begin{eqnarray}
\rm \frac{\dot m_3}{Nm_3}=\frac{\ell_3}{\eta^3},\label{m3} \\
\rm \frac{\dot m_1}{Nm_1}=\frac{\ell_1}{\eta^3}\label{m1} ,
\end{eqnarray}
the constants being $\rm \ell_3=\frac{\ell_{32}-\ell_{13}}{3}$ and
$\rm \ell_1=\frac{\ell_{13}-\ell_{21}}{3}$, also, these constants
satisfy $\sum_{j=1}^3 \ell_j=0$. Now that equations
(\ref{m2})-(\ref{m1}) are written in a more manageable way,
obtaining the solutions is straightforward, which are given by
\begin{equation}
\rm m_i(t)=\alpha_i Exp\left[\ell_i \int \frac{Ndt}{\eta^3}\right],
\end{equation}
where $\rm \Pi_{j=1}^3\alpha_j=1$.

\section{Some exact solutions in the gauge N=1}

\subsection{Volume for $\gamma\neq1$}
Using equation (\ref{00-alfa}) along with the relations $\rm
\frac{\dot m_i}{m_i}=\ell_i/\eta^3$ we get
\begin{eqnarray}
&&\rm
\frac{\dot m_1}{m_1} \frac{\dot m_2}{m_2}+\frac{\dot m_2}{m_2}\frac{\dot m_3}{m_3}+\frac{\dot m_1}{m_1}\frac{\dot m_3}{m_3}+3 \left(\frac{\dot \eta}{\eta}\right)^2-8\pi G \alpha_\phi \rho = \nonumber\\
&&\rm \frac{(\ell_1 \ell_2 +\ell_1 \ell_3 + \ell_2
\ell_3)}{\eta^6}+3 \left(\frac{\dot \eta}{\eta}\right)^2-8\pi
G\alpha_\phi \rho_\gamma \eta^{-3(1+\gamma)}=0,
\end{eqnarray}
however, we can obtain a master equation for $\eta$, which is given
by
\begin{equation}
\rm-\frac{\ell^2}{\eta^6}+3 \left(\frac{\dot
\eta}{\eta}\right)^2-8\pi G \alpha_\phi\rho_\gamma
\eta^{-3(1+\gamma)}=0,
\end{equation}
where we have identified $\rm (\ell_1 \ell_2 +\ell_1 \ell_3 + \ell_2
\ell_3)=-(1/2)\sum_{j=1}^3\ell_j^2=-\ell^2$ using the constraint
$\sum_{j=1}^3 \ell_j=0$.

The previous equation can be expressed as
\begin{equation}
\rm \frac{\eta^2d\eta}{\sqrt{\frac{8\pi G \alpha_\phi \rho_\gamma
\eta^{3(1-\gamma)} + \ell^2}{3}}}=dt,\label{master}
\end{equation}
whose general solutions are given in terms of a hypergeometric
function,
\begin{equation}
\rm  \Delta t =\frac{\eta^3}{3\ell} \, _2F_1\left[\frac{1}{2},
\frac{1}{1-\gamma},\frac{2-\gamma}{1-\gamma}, -\frac{8\pi G
\alpha_\phi \rho_\gamma}{\ell^2}\eta^{3(1-\gamma)}
\right],\quad(\gamma\neq1).
\end{equation}

\subsection {Stiff matter scenario ($\gamma=1$) without scalar field ($\alpha_\phi=1$)}
When the scenario of stiff matter, $\gamma=1$, is considered, the
master equation (\ref{master}) reduces to
\begin{equation}
\rm \frac{\eta^2 d\eta}{\sqrt{\frac{8\pi G \alpha_\phi \rho_1}{3} +
\frac{\ell^2}{3}}}=dt,
\end{equation}
which has a solution of the form
\begin{equation}
\rm \eta^3=\left[ \eta_0 +3 c_0 \Delta t\right],
\end{equation}
where $\rm c_0=\sqrt{\frac{8\pi G \alpha_\phi \rho_1}{3} +
\frac{\ell^2}{3}}$ has dimensions of (1/s), and we can see that the
isotropic volume function depends linearly with respect to time. For
the stiff matter case the functions $\rm m_i(t)$ will be given by
\begin{equation}
\rm m_i(t)=\alpha_i \left    [ \eta_0+3c_0 \Delta t
\right]^{\frac{\ell_i}{3c_0}},
\end{equation}
where we can clearly notice that dynamically we do not obtain
isotropization. In passing, we note that, even when the scalar field
is not considered, this type of solution could be related to the so
called K-essence paradigm, since a simple class of purely kinetic
K-essence cosmological models behave like a stiff fluid
\citep{Socorro et al:2010, Socorro et al:2014, Espinoza et al:2014}.

\subsection{Dust scenario ($\gamma=0$)}
In this case, we also present the solution for the scalar field, via
equation (\ref{scalar-field}). For the dust scenario, $\gamma=0$,
the master equation (\ref{master}) takes the following form
\begin{equation}
\rm \frac{\eta^2 d\eta}{\sqrt{\frac{8\pi G \alpha_\phi \rho_0 \eta^3
+ \ell^2}{3}}}=dt,
\end{equation}
where the solution is given by
\begin{equation}
\rm \frac{4\sqrt{3}\pi G \alpha_\phi \rho_0}{\ell} \Delta t=
\sqrt{\frac{8\pi G \alpha_\phi \rho_0}{\ell^2} \eta^3 + 1} - b_2,
\end{equation}
here $\rm b_2$ is a dimensionless constant related to the initial
conditions for the dust era. From the last equation we can obtain
$\rm \eta^3$, which reads
\begin{equation}
\rm \eta^3=b_0 \left[\left( b_1 \Delta t + b_2\right)^2 - 1 \right],
\label{volume}
\end{equation}
where $\rm b_0=\ell^2/(8\pi G \alpha_\phi \rho_0)$ is dimensionless
implying that the constant $\ell$ have dimension of (1/s), similarly
$\rm b_1=(4\sqrt{3}\pi G \alpha_\phi \rho_0)/\ell$ has the same
dimensions, which in turn enables us to find that the expression for
$\rm m_i(t)$, which is given by
\begin{equation}
\rm m_i(t)=\alpha_i \left    ( 1- \frac{2}{b_1 t + b_2+1}
\right)^{q_i}, \label{anisotropias}
\end{equation}
where $\rm q_i=\frac{\sqrt{3}\ell_i}{3\ell}$ with $\rm
q_2^2+q_3^2+q_2q_3=\frac{1}{3}$. It is important to emphasize that
the anisotropic parameter acquires a constant value in the limit
$t\to\infty$, signaling that this anisotropic model eventually
transits to the flat FRW model, which is the model that we ``see''
today. Additional arguments supporting this claim will be given in
the next section.

Following \citep{Tripathy et al:2012}, we consider the volume
deceleration parameter,
\begin{equation}
\rm q(t)=-\frac{v\ddot{v}}{\dot v^2},
\end{equation}
where $\rm v=\eta^3=a_1a_2a_3$ is the (isotropic) volume function of
the Bianchi type I model, which is in this case given by the exact
solution (\ref{volume}). Explicitly, we have,
\begin{equation}
\rm q(t)=-\frac{1}{2}\left(1-\frac{1}{(b_1t+b_2)^2}\right),
\label{q}
\end{equation}
from which we deduce $\rm\lim\limits_{t\to\infty}q(t)=-\frac{1}{2}$.
These observations indicate that the universe presents a volume
accelerated expansion in the dust epoch.

\begin{figure}[ht]
\begin {center}
\includegraphics[width=0.5\textwidth]{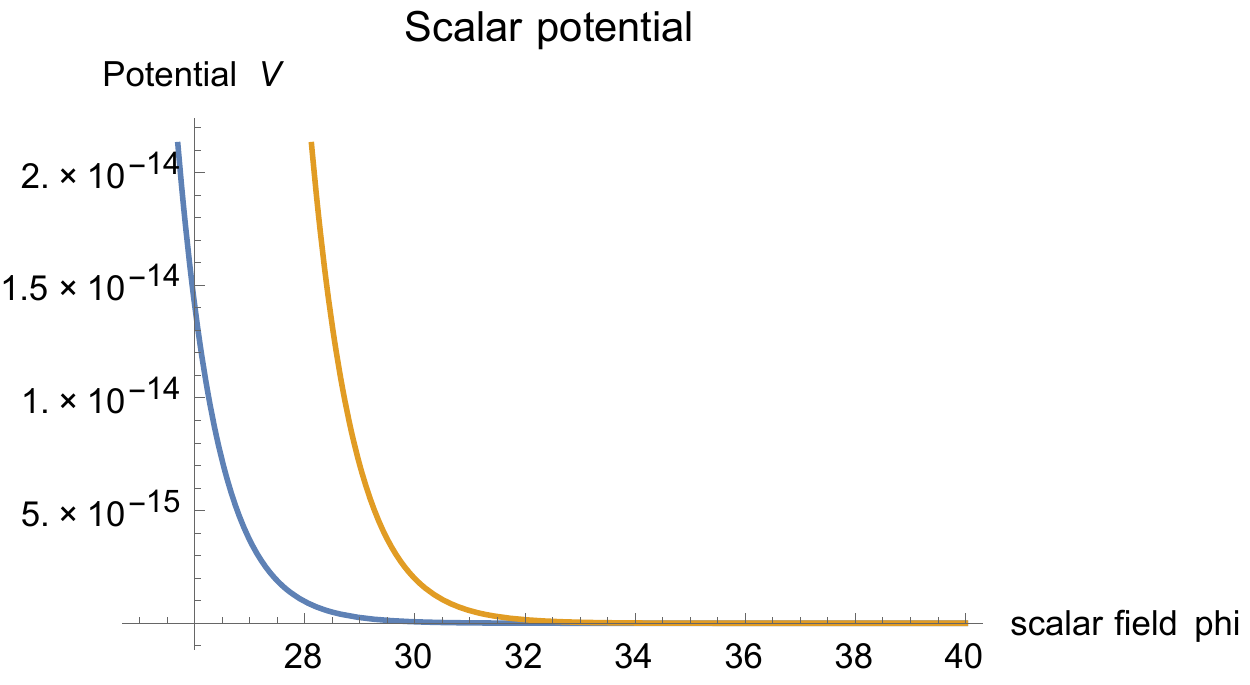}
\caption{Behavior of the scalar potential as a function of the
scalar field, given by Eq. (\ref{pot}). The blue solid line
corresponds to $\rm m_\phi=5$ (dark energy), while the yellow solid
line corresponds to $\rm m_\phi=14$ (dark matter).}
\label{fig:vol_phi}
\end{center}
\end{figure}

Now, the scalar field can be determined from (\ref{scalar-field}),
\begin{equation}
\rm \Delta \phi=\int \frac{dt}{\sqrt{b_0}\sqrt{(b_1 t + b_2)^2-1}},
\end{equation}
making the following change of variables $\rm u=b_1 t+b_2$ we get
$\rm \Delta \phi=(1/b_1\sqrt{b_0}) \, Ln \left[
u+\sqrt{u^2-1}\right]$, to finally obtain $\rm u=cosh\left[
\frac{1}{2}\sqrt{\frac{3}{2}(1+\frac{1}{m_\phi})} \Delta
\phi\right]$. Substituting in (\ref{potential}) we find
\begin{eqnarray}
&&\rm V(t)=8\pi G \rho_0 m_\phi  \frac{1}{ \left( b_1 \Delta t + b_2\right)^2 - 1}, \quad \leftrightarrow \nonumber\\
&&\rm  V(\phi)=8\pi G \rho_0 m_\phi \,
Csch^2\left(\frac{1}{2}\sqrt{\frac{3}{2}\left(1+\frac{1}{m_\phi}\right)}
\Delta \phi \right).  \label{pot}
\end{eqnarray}
At this point, it is worth mentioning that this kind of potential is
consistent with a (volume) acelerated expansion in the dust
scenario. The behavior of the potential in terms of the scalar field
is shown in Fig.~(\ref{fig:vol_phi}).

\begin{figure}[ht]
\begin {center}
\includegraphics[width=0.5\textwidth]{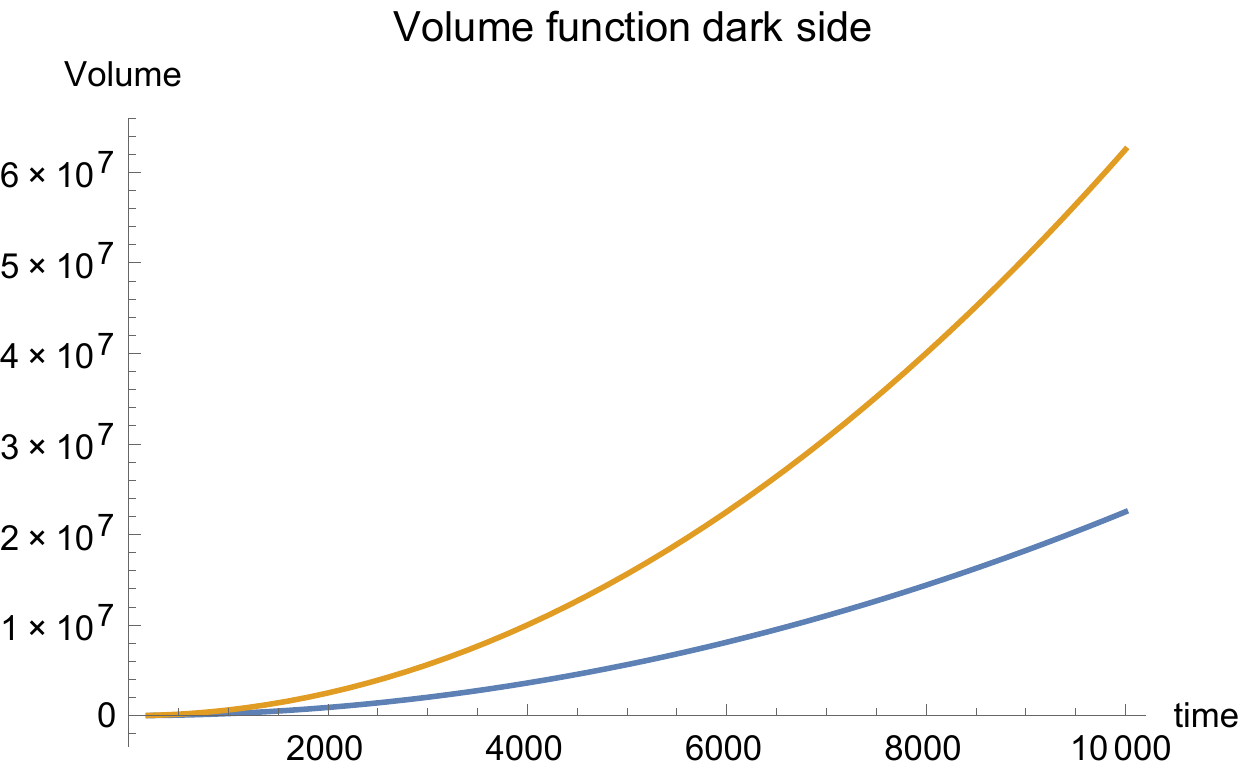}
\caption{Behavior of the isotropic parameter (volume function) in
the dust scenario, given by equation (\ref{volume}). Where we have
set $\rm b_1=10, b_2=10, b_0=0.001$. The yellow solid line
corresponds to $\rm m_\phi=14$ and the solid blue line corresponds
to $\rm m_\phi=5$.} \label{fig:vol_t}
\end{center}
\end{figure}

On the other hand, the cosmological term for this scenario is
\begin{equation}
\rm \Lambda(t)=\frac{4\pi G \rho_0 m_\phi}{ \left[ \left( b_1 \Delta
t + b_2\right)^2 -1\right]}
\end{equation}
which is a behavior that anisotropic universes exhibit. We can
observe that the universe has a bigger expansion rate for $\rm
m_\phi=14$ (which accounts for approximately $\rm 70\%$ in the total
matter density for dust scenario) than for $\rm m_\phi=5$. The
behavior of these functions can be seen in Fig.~(\ref{fig:vol_t})

The time evolution for the average volume function in this scenario
will be given by
\begin{equation}
\rm \eta^3=b_0 \left[\left( b_1 \Delta t + b_2\right)^2 - 1 \right],
\label{volume1}
\end{equation}
where $\rm b_0=\ell^2/(8\pi G \alpha_\phi \rho_0)$ and $\rm
b_1=(4\sqrt{3}\pi G \alpha_\phi \rho_0)/\ell$, that in terms of the
scalar field can be written as
\begin{equation}
\rm  \eta^3=b_0 sinh^2 \left[
\frac{1}{2}\sqrt{\frac{3}{2}\left(1+\frac{1}{m_\phi}\right)} \Delta
\phi\right],
\end{equation}
where we have used $$\rm u=b_1 t + b_2=
cosh\left[\frac{1}{2}\sqrt{\frac{3}{2}\left(1+\frac{1}{m_\phi}\right)}
\Delta \phi\right].$$ From the previous equation we can infer that
the volume function has a stronger dependence on the dark scenario.
We can see that for early times the value of $\rm m_\phi$ is small
having a growing behavior; but when $\rm m_\phi$ has large values
the volume function has decelerate behavior.

\subsection{Anisotropic parameters}
In anisotropic cosmology, the Hubble parameter $\rm H$ is defined in
analogy with the FRW cosmology:
\begin{equation}
\rm H=\frac{\dot a}{a}=\frac{\dot \eta }{\eta
}=\frac{1}{3}\left(H_x+H_y+H_z \right),
\end{equation}
where $\rm H_x=\frac{\dot a_1}{a_1}$, $\rm H_y=\frac{\dot
a_2}{a_2}$, and $\rm H_z=\frac{\dot a_3}{a_3}$.

The scalar expansion $\rm \theta$, shear scalar $\rm \sigma^2$ and
the average anisotropic parameter $\rm \overline A_m$ are defined as
\begin{equation}
\rm \theta=\sum_{i=1}^3 \frac{\dot a_i}{a_i}= 3H, \qquad
\sigma^2=\frac{1}{2} \left(\sum_{i=1}^3 H_i^2 - \frac{1}{3}\theta^2
\right), \qquad \overline A_m = \rm \frac{1}{3} \sum_{i=1}^3 \left(
\frac{H_i-H}{H}\right)^2, \label{aniso-parameter}\end{equation}
respectively.

Using the results for the average scale factor $\eta$ and the
dimensionless anisotropic functions $\rm m_i$, the average
anisotropic parameter is $\rm \overline
A_m=\frac{1}{3}\frac{\sum_{i=1}^3 \ell_i^2}{\eta^6 \left(\frac{\dot
\eta}{\eta}\right)^2}=\frac{3\sum_{i=1}^3 \ell_i^2}{4b_0^2 b_1^2
\,u^2}$, which can be written in terms of the scalar field using the
variable u defined previously,
\begin{equation}
\rm \overline A_m=\frac{3\sum_{i=1}^3 \ell_i^2}{4b_0^2 b_1^2} \,
Sech^2\left[\frac{1}{2}\sqrt{\frac{3}{2}\left(1+\frac{1}{m_\phi}\right)}
\Delta \phi\right].\label{aniso-parameter2}
\end{equation}
This parameter goes to zero for $\phi \to \infty$, causing the
isotropization of the model. The other parameters aquire the form
\begin{eqnarray}
\rm \theta &=& \rm 3\frac{\dot \eta}{\eta}=2 b_1
\frac{u}{u^2-1},\label{expansion}\\
\rm \sigma^2&=& \rm \frac{1}{2} \left( \frac{\dot
m_i}{m_i}\right)^2=\frac{1}{2}\frac{\sum_{i=1}^3 \ell_i^2}{ \eta^6}
=\frac{1}{2}\frac{\sum_{i=1}^3 \ell_i^2}{ b_0^2 (u^2-1)^2},
\label{shear}
\end{eqnarray}
Following \citep{Pradhan2011} and references therein, where the
authors precise that the red-shift studies place the limit
$\frac{\sigma}{\theta} \leq 0.3$ on the ratio of shear $\sigma$ to
Hubble constant H in the neighborhood of our Galaxy today in order
to have a sufficiently isotropic cosmological model, we obtain,
\begin{equation}
\rm \frac{\sigma^2}{\theta^2}= \frac{ \sum_{i=1}^3 \ell_i^2}{8b_0^2
b_1^2}\,\frac{1}{u^2}=\frac{ \sum_{i=1}^3 \ell_i^2}{8b_0^2 b_1^2}
Sech^2\left[
\frac{1}{2}\sqrt{\frac{3}{2}\left(1+\frac{1}{m_\phi}\right)} \Delta
\phi\right]. \label{reason}
 \end{equation}
Hence, we must have, $\rm \Delta \phi \leq
\frac{2}{\sqrt{\frac{3}{2}\left(1+\frac{1}{m_\phi}\right)}}
ArcSech\left( \frac{6\sqrt{2} b_0 b_1}{10\sqrt{\sum_{i=1}^3
\ell_i^2}}\right)$, where we consider that the constants should be
fixed by the observational data, following the procedure in
reference \citep{Goswami et al:2020}. Using this result, the average
anisotropic parameter is $\rm \overline A_m \leq 0.54$, see
equations (\ref{aniso-parameter}) and (\ref{reason}).

\section{Conclusions and remarks}

In this work we have characterized a sufficiently isotropic universe
presenting a volume accelerated expansion in a dust stage,
introducing a combination of results using two different approaches
within an anisotropic cosmological model.

Employing a Misner-like transformation we can consider a
decomposition into isotropic and anisotropic parts, where the
properties of the latter are preserved, appearing as constraint
equations (see (\ref{mi})). Also, we extend the identification $\rm
V(\phi)=2\Lambda$, (with $\Lambda$ a \textit{constant}), to the
dynamical case $\rm V(\phi(t))=2\Lambda(t)$.

The other idea was to consider a law between the energy density of
the scalar field and that of the ordinary matter as follows: $\rm
\rho_\phi= m_\phi \rho$; which results as a consequence of
considering barotropic equations of state, $\rm P_\phi=\omega_\phi
\rho_\phi$ and $\rm p=\gamma \rho$, and the equality of the
corresponding barotropic indices. We found that for the solutions to
the EKG equations to be consistent, the barotropic parameters must
be equal.

We were able to find analytical solutions for the gravitational
potentials and for the scalar field in a dust scenario, using the
scaling solutions (previously found) between the energy density of
the scalar field and that of the standard matter. It is worth
mentioning that, in a dust scenario, the gravitational potentials
have the same structure independently of wether or not the scalar
field is taken into account: $\alpha_\phi=1$ without scalar field;
$\rm\alpha_\phi=1+m_\phi$ with scalar field. We remark that
parameter $\rm m_\phi$ is responsible for the description of the
dark side of the universe, where $\rm m_\phi=5$ accounts for dark
matter and $\rm m_\phi=14$ for dark energy. Considering this two
values for the evolution of the universe, it turns out that for $\rm
m_\phi=14$ the growth of the volume is faster than for $\rm
m_\phi=5$, as expected, as shown in Fig.~(\ref{fig:vol_t}). The
exact solution for the isotropic volume shows that the volume of the
universe suffers an accelerated growth (in the dust scenario). Also,
we found that (for the dust stage) the anisotropic Bianchi type I
cosmological model evolves into the isotropic flat FRW model. This
is supported by the fact that the anistropic parameters $\rm m_i$
acquire constant values for $t\to\infty$. Furthermore, the average
anisotropy parameter $\rm\overline{A}_m$ is asymptotically null.
Additionally, by considering the bound
$\frac{\sigma}{\theta}\leq0.3$, parameter $\rm\overline{A}_m$ was
constrained to $\rm\overline{A}_m\leq0.54$, which is consistent with
the form considered in \citep{Amirhashchi:2018}.

Finally, we remark that our findings indicate (regarding our toy
model) that for the universe to present a volume accelerated
expansion today, considering the volume deceleration parameter
(\ref{q}) \citep{Tripathy et al:2012}, and the scalar potential must
behaves as a hyperbolic cosecant, as shown in Fig.
(\ref{fig:vol_phi}), implying that $\Lambda(t)\sim 1/t^2$. We also
would like to note that we could not find this type of potential in
the literature regarding studies in scalar field anisotropic
cosmology.
\\

\acknowledgments{ \noindent This work was partially supported by
PROMEP grants UGTO-CA-3. S.P.P. and J.S. were partially supported
SNI-CONACYT. This work is part of the collaboration within the
Instituto Avanzado de Cosmolog\'{\i}a and Red PROMEP: Gravitation
and Mathematical Physics under project {\it Quantum aspects of
gravity in cosmological models, phenomenology and geometry of
space-time}. Many calculations where done by Symbolic Program REDUCE
3.8.}


\begin{thebibliography}{99}
\bibitem[\protect\citeauthoryear{Abdel}{1990}]{Abdel:1990}  Abdel A.M.M.:
 \emph{Gen. Rel. Grav.} {\bf 22}, 655 (1990).
\bibitem[\protect\citeauthoryear{{Arbab} \& {Abdel}}{1994}]{Abdel:1994} Arbab, A.I., \&  Abdel-Rahaman, A.M.M.:
 \emph{Phys. Rev. D} {\bf 50}, 7725 (1994).
\bibitem[\protect\citeauthoryear{Akarsu et al.}{2019}]{Akarsu et al:2019} Akarsu, \"O., et al.:
    \emph{Phys. Rev. D} {\bf 100}, 023532 (2019).
\bibitem[\protect\citeauthoryear{Amirhashchi}{2018}]{Amirhashchi:2018} Amirhashchi, H.:  \emph{Phys. Rev. D} {\bf 97}, 063515 (2018).
\bibitem[\protect\citeauthoryear{Amirhashchi}{2019}]{Amirhashchi:2019} Amirhashchi, H. \& Amirhashchi, S.:
    \emph{Phys. Rev. D} {\bf 99}, 023516 (2019).

\bibitem[\protect\citeauthoryear{Arbad}{1997}]{Arbab:1997} Arbab, A.I.: \emph{Gen. Rel. Grav.} {\bf 29}, 61 (1997).
\bibitem[\protect\citeauthoryear{Arbad}{2001}]{Arbab2001} Arbab, A.I.: \emph{Spacetime and substance} {\bf 1}, 39 (2001).
\bibitem[\protect\citeauthoryear{Arbad}{2003}]{Arbab2003} Arbab, A.I.: \emph{Class. Quantum Grav.} {\bf 20}, 93 (2003).
\bibitem[\protect\citeauthoryear{Arbad}{2004}]{Arbab2004} Arbab, A.I.: \emph{Astrophys. Space Sci.} {\bf 291}, 141 (2004).
\bibitem[\protect\citeauthoryear{Aroonkumar}{1993}]{Aroonkumar:1993} Aroonkumar Beesham.: \emph{Phys. Rev. D} {\bf 48}, 3539 (1993).
\bibitem[\protect\citeauthoryear{Aroonkumar}{1994}]{Aroonkumar:1994} Aroonkumar Beesham.: \emph{Gen. Rel. Grav.} {\bf 26}, 159 (1994).
\bibitem[\protect\citeauthoryear{{Bal} \& {Singh}}{2008}]{Bal:2008} Bal, R., \& Singh, J.P.:
     \emph{Int. J. of Theor. Phys.} {\bf 47}, 3288 (2008).
\bibitem[\protect\citeauthoryear{{Bali} \& {Jain}}{2002}]{Bali:2002} Bali, R., \& Jain, V.C.: \emph{Pramana J. Phys.} {\bf 59}, 1 (2002).
\bibitem[\protect\citeauthoryear{Bali}{2011}]{Bali:2011} Bali, R.: \emph{Int. J. of Theor. Phys.} {\bf 50}, 3043 (2011).
\bibitem[\protect\citeauthoryear{Belinchon}{2008}]{Belinchon:2008} Belinch\'on, J.A.: \emph{Int. J. Mod. Phys. A} {\bf 23}, 5021 (2008).
\bibitem[\protect\citeauthoryear{{Belinskii} \& {Khalatnikov}}{1972}]{Belinskii:1972} Belinskii, V.A., \& Khalatnikov, I.M.: \emph{Sov. Phys. JETP} {\bf 63}, 1121 (1972).
\bibitem[\protect\citeauthoryear{{Birkel} \& {Sarkar}}{1997}]{Birkel:1997} Birkel, M., \& Sarkar, S.:
    \emph{Astropart. Phys.} {\bf 6}, 197 (1997).
\bibitem[\protect\citeauthoryear{Burd}{1988}]{Burd:1988} Burd, A.B., \& Barrow, J.D., \emph{Nucl. Phys. B} {\bf 308}, 929 (1988).
\bibitem[\protect\citeauthoryear{{Bylan} \& {Scialom}}{1998}]{Bylan:1998} Bylan, S.,\& Scialom, D.: \emph{Phys. Rev. D} {\bf 57}, 6065 (1998).
\bibitem[\protect\citeauthoryear{{Carneiro} \&  {Lima}}{2005}]{Carneiro:2005} Carneiro, S., \&  Lima, J.A.S.:
    \emph{Int. J. Mod. Phys. A} {\bf 20}, 2465 (2005).
\bibitem[\protect\citeauthoryear{Carvalho et al.}{1992}]{Carvalho et al:1992} Carvalho, J.C. et al.:
    \emph{Phys. Rev. D} {\bf 46}, 2404 (1992).
\bibitem[\protect\citeauthoryear{Copeland et al.}{1998}]{Copeland et al:1998} Copeland, E.J. et al.:
    \emph{Phys. Rev. D} {\bf 57}, 4686 (1998).
\bibitem[\protect\citeauthoryear{{Cunha} \&  {Santos}}{2004}]{Cunha:2004} Cunha, J.V., \&  Santos, R.C.:
    \emph{Int. J. Mod. Phys. D} {\bf 13}, 1321 (2004).
\bibitem[\protect\citeauthoryear{{Chen} \& {Wu}}{1990}]{Chen:1990} Chen, W., \&  Wu, Y.S.:
    \emph{Phys. Rev. D} {\bf 41},695 (1990).
\bibitem[\protect\citeauthoryear{Espinoza et al}{2014}]{Espinoza et al:2014}  Espinoza Garc\'ia, Abraham., et al.:
    \emph{Int. J. of Theor. Phys.} {\bf 53} (9), 3066-3077 (2014)
\bibitem[\protect\citeauthoryear{Esposito et al.}{2007}]{Esposito et al:2007} Esposito, G., et al.:
    \emph{Class. Quantum Grav.} {\bf 24}, 6255 (2007).
\bibitem[\protect\citeauthoryear{{Ferreira} \& {Joyce}}{1997}]{Ferreira:1997} Ferreira, P.G., \& Joyce, M., \emph{Phys. Rev. Lett.} {\bf 79}, 4740 (1997).
\bibitem[\protect\citeauthoryear{{Ferreira} \& {Joyce}}{1998}]{Joyce:1998} Ferreira, P.G., \& Joyce, M.:
    \emph{Phys. Rev. D} {\bf 58}, 023503 (1998).
\bibitem[\protect\citeauthoryear{{Folomeev} \& {Gurovich}}{2000}]{Folomeev:2000} Folomeev, V.N., \&  Gurovich, V. Ts.:
    \emph{Gen. Rel. Grav.} {\bf 32}(7), 1255 (2000).
\bibitem[\protect\citeauthoryear{Fomin et al.}{2005}]{Fomin:2005} Fomin, P.I., et al.: \emph{preprint} [gr-qc/0509042].
\bibitem[\protect\citeauthoryear{Goswami et al.}{2020}]{Goswami et al:2020} Goswami, G.K., et al.: \emph{Mod. Phys. Let. A}
    2050086 (2020), https://doi.org/10.1142/S0217732320500868.
\bibitem[\protect\citeauthoryear{Halliwell}{1985}]{Halliwell:1985} Halliwell, J.:
    \emph{Phys. Lett. B} {\bf 185}, 341 (1985).
\bibitem[\protect\citeauthoryear{{Jamil} \& {Debnath}}{2011}]{Jamil:2011} Jamil, M., \& Debnath, U.:
    \emph{Int. J. of Theor. Phys.} {\bf 50}, 1602 (2011).
\bibitem[\protect\citeauthoryear{Kalligas et al.}{1992}]{Kalligas et al:1992}  Kalligas, D. et al.:
    \emph{Gen. Rel. Grav.} {\bf 24}, 351 (1992).
\bibitem[\protect\citeauthoryear{Knop et al.}{2003}]{Knop et al:2003} Knop R.A. et al.: \emph{Astrophys. J.} {\bf 598}, 102 (2003).
\bibitem[\protect\citeauthoryear{{Kumar} \& {Singh}}{2007}]{Kumar:2007} Kumar, S., Singh, C.P.: \emph{ Astrophys
    Space Sci} {\bf 312}, 57 (2007).
\bibitem[\protect\citeauthoryear{{Liddle} \& {Sharrer}}{1998}]{Liddle:1998} Liddle, A.R., \& Sharrer, R.J.:
     \emph{Phys. Rev. D} {\bf 59}, 023509 (1998).
\bibitem[\protect\citeauthoryear{{Lucchin} \& {Matarrese}}{1985}]{Lucchin:1985} Lucchin, F., \& Matarrese, S.:
    \emph{Phys. Rev. D} {\bf 32}, 1316 (1985).
\bibitem[\protect\citeauthoryear{{Lima} \& {Maia}}{1994}]{Maia:1994} Lima, J.A.S., \&  Maia, J.M.F.:
    \emph{Phys. Rev. D} {\bf 49}, 5597 (1994).
\bibitem[\protect\citeauthoryear{{Lima} \& {Carvalho}}{1994}]{Lima:1994} Lima, J.A.S., \& Carvalho, J.C.:
    \emph{Gen. Rel. Grav.} {\bf 26}, 909 (1994).
\bibitem[\protect\citeauthoryear{{Lima} \& {Trodden}}{1996}]{Trodden:1996} Lima, J.A.S., \& Trodden, M.:
     \emph{Phys. Rev. D} {\bf 53}, 4280 (1996).
\bibitem[\protect\citeauthoryear{{Martinez} \& {Sanz}}{1995}]{Martinez:1995} Martinez-Gonzalez, E., \& Sanz, J.L.:
    \emph{Astronomy and Astrophysics} {\bf 300}, 346 (1995).
\bibitem[\protect\citeauthoryear{Mukhopadhyay et al.}{2011}]{Mukhopadhyay:2011} Mukhopadhyay, U. et al.:
    \emph{Int. J. of Theor. Phys.} {\bf 50}, 752 (2011).
\bibitem[\protect\citeauthoryear{{Overduin} \& {Cooperstock}}{1998}]{Overduin:1998} Overduin, J.M.,  \& Cooperstock,  F.I.: \emph{Phys. Rev. D} {\bf 58}, 043506 (1998).
\bibitem[\protect\citeauthoryear{Pavon}{1991}]{Pavon:1991} Pavon, D.: \emph{Phys. Rev. D} {bf 43}, 375 (1991).
\bibitem[\protect\citeauthoryear{Perlmutter et al.}{1999}]{Perlmutter et al:1999} Perlmutter, S. et al.:
    \emph{astrophys. J.} {\bf 517}, 565 (1999).
\bibitem[\protect\citeauthoryear{{Pradhan} \& {Kumar}}{2001}]{Kumar:2001} Pradhan, A. \&  Kumar, A.:
    \emph{Int. J. of Mod. Phys. D} {\bf 10}, 291 (2001).
\bibitem[\protect\citeauthoryear{Pradhan}{2003}]{Pradhan:2003} Pradhan, A.: \emph{Int. J. of Mod. Phys. D} {\bf 12}, 941 (2003).
\bibitem[\protect\citeauthoryear{Pradhan}{2007}]{Pradhan:2007} Pradhan, A.: \emph{Fizika B} {\bf 16}, 205 (2007).
\bibitem[\protect\citeauthoryear{Pradhan}{2009}]{Pradhan:2009} Pradhan, A.: \emph{Commun. Theor. Phys.} {\bf 51}, 367 (2009).
\bibitem[\protect\citeauthoryear{{Pradhan} \&  {Pandey}}{2003}]{Pandey:2003} Pradhan, A., \& Pandey, A.P.:
    \emph{Int. J. of Mod. Phys. D} {\bf 12}, 1299 (2003).
\bibitem[\protect\citeauthoryear{{Pradhan} \&  {Pandey}}{2006}]{PradhanPandey:2006} Pradhan, A., \&  Pandey, A.P.:
    \emph{Astrophys. and Spa. Sci.} {\bf 301}, 127 (2006).
\bibitem[\protect\citeauthoryear{Pradhan et al.}{2007}]{Pradhan2007} Pradhan, A. et al.:
    \emph{Int. J. of Theor. Phys.} {\bf 46}, 2774 (2007).
\bibitem[\protect\citeauthoryear{Pradhan et al.}{2007}]{Pradhana2007}  Pradhan, A. et al.:
    \emph{Rom. J. Phys.} {\bf 52}, 445 (2007).
\bibitem[\protect\citeauthoryear{Pradhan et al.}{2008}]{Pradhan2008} Pradhan, A. et al.:
    \emph{Brazilian J. of Phys.} {\bf 38}, 167 (2008).
\bibitem[\protect\citeauthoryear{Pradhan et al.}{2011}]{Pradhan2011} Pradhan, A. et al.:
    \emph{Int. J. Theor. Phys.} {\bf 50}, 2923 (2011).
\bibitem[\protect\citeauthoryear{Pradhan et al.}{2012}]{Pradhan2012} Pradhan, A. et al.:
     \emph{Astrophys. Space Sci.} {\bf 337}, 401 (2012).
\bibitem[\protect\citeauthoryear{Pradhan et al.}{2015}]{Pradhan2015} Pradhan, A. et al.:
    \emph{Indian J. Phys.}  {\bf 89}(5), 503 (2015).
\bibitem[\protect\citeauthoryear{{Rahman} \& {Ansary}}{2013}]{Ansary:2013} Rahman, M.A., \& Ansary, M.:
    \emph{Prespace time J.} {\bf 4}, 871 (2013).
\bibitem[\protect\citeauthoryear{Ray et al.}{2011}]{Ray et al:2011} Ray, S. et al.:
    \emph{Int. J. of Theor. Phys.} {\bf 50}, 939 (2011).
\bibitem[\protect\citeauthoryear{Riess et al.}{1998}]{Riess et al:1998} Riess, A.G., et al., \emph{astron. J.} {\bf 116}, 1009 (1998).
\bibitem[\protect\citeauthoryear{Riess et al.}{2004}]{Riess et al:2004} Riess, A.G., et al.:
    \emph{Astrophys. J.} {\bf 607}, 665 (2004).
\bibitem[\protect\citeauthoryear{Saha}{2006}]{Saha:2006} Saha, B.: \emph{Astrophys. and Spa. Sci.} {\bf 302}, 83  (2006).
\bibitem[\protect\citeauthoryear{Shahalam et al.}{2015}]{Shahalam et al:2015} Shahalam, M. et al.:
    \emph{Eur. Phys. J. C} {\bf 75}, 395 (2015).
\bibitem[\protect\citeauthoryear{Shen}{2013}]{Shen:2013} Shen, M.: \emph{Int. J. of Theor. Phys.} {\bf 52}, 178 (2013).
\bibitem[\protect\citeauthoryear{{Silveira} \& {Waga}}{1997}]{Silveira:1997} Silveira V., \& Waga, I.:
    \emph{Phys. Rev. D} {\bf 56}, 4625 (1997).
\bibitem[\protect\citeauthoryear{Singh et al.}{1998}]{Singh:1998} Singh, T. et al.: \emph{Gen. Rel. Grav.} {\bf 30}, 573 (1998).
\bibitem[\protect\citeauthoryear{Singh et al.}{2008}]{Singh et al:2008} Singh, J.P. et al.: \emph{Astrophys. Spa. Sci.} {\bf 314}, 83 (2008).
\bibitem[\protect\citeauthoryear{Singh}{2008}]{Singh:2008} Singh, J.P., \emph{Astrophys Spa. Sci.} {\bf 318}, 103 (2008).
\bibitem[\protect\citeauthoryear{Singh et al.}{2013}]{Singh et al:2013} Singh, M.K. et al.:
    \emph{Int. J. of Phys.} {\bf 1}, 77 (2013).
\bibitem[\protect\citeauthoryear{Socorro et al.}{2010}]{Socorro et al:2010} Socorro, J., et al.:
    \emph{Rev. Mex. F\'is.} {\bf 56}(2), 166-171 (2010).
\bibitem[\protect\citeauthoryear{Socorro et al.}{2014}]{Socorro et al:2014} Socorro, J., et al.:
    \emph{Advances in High Energy Phys.} 805164 (2014).
\bibitem[\protect\citeauthoryear{Socorro et al.}{2015}]{Socorro et al:2015} Socorro, J., et al.:
    \emph{Astrophys. Spa. Sci.} {\bf 360}, 20 (2015).
\bibitem[\protect\citeauthoryear{{Sola} \& {Stefancic}}{2005}]{Stefancic:2005}  Sola, J., \& Stefancic, H.:
    \emph{Phys. Lett. B} {\bf 624}, 147 (2005).
\bibitem[\protect\citeauthoryear{{Sola} \& {Stefancic}}{2006}]{Sola:2006} Sola, J., \& Stefancic, H.:
    \emph{Mod. Phys. Lett. A} {\bf 21}, 479 (2006).
\bibitem[\protect\citeauthoryear{Spergel et al.}{2007}]{Spergel et al:2007} Spergel, D.N., et al.:
    \emph{Astrophys. J. Suppl.} {\bf 170}, 377 (2007).
\bibitem[\protect\citeauthoryear{Starobinsky}{1998}]{Starobinsky:1998} Starobinsky, A.A., \emph{JETP Letters} {\bf 8}, 757 (1998).
\bibitem[\protect\citeauthoryear{Tegmark et al.}{2004}]{Tegmark et al:2004} Tegmark, M., et al.: \emph{Phys. Rev. D} {\bf 69}, 103501 (2004).
\bibitem[\protect\citeauthoryear{Tripathy et al.}{2012}]{Tripathy et al:2012} Tripathy, S.K., et al.: \emph{Astrophys. Spa. Sci.} {\bf 340}, 211 (2012).
\bibitem[\protect\citeauthoryear{Tripathy}{2013}]{Tripathy:2013} Tripathy, S.K., \emph{Int. J. Theor. Phys.} {\bf 52}, 4218 (2013).
\bibitem[\protect\citeauthoryear{Vishwakarma}{2000}]{Vishwakarma:2000} Vishwakarma, R.G.: \emph{Class Quantum Grav.} {\bf 17}, 3833 (2000).
\bibitem[\protect\citeauthoryear{Vishwakarma}{2001}]{Vishwakarma:2001} Vishwakarma, R.G.: \emph{Gen. Rel. Grav.} {\bf 33}, 1973 (2001).
\bibitem[\protect\citeauthoryear{Wand et al.}{1993}]{Wand et al:1993} Wand, D. et al.. \emph{Ann (NY) Acad. Sci.} {\bf 688}, 647 (1993).
\bibitem[\protect\citeauthoryear{Weetterich}{1998}]{Weetterich:1998} Weetterich, C.: \emph{Nucl. Phys. B} {\bf 302}, 668 (1998).
\bibitem[\protect\citeauthoryear{Zia et al.}{2019}]{Zia et al:2019} Zia, R., et al: \emph{New Astronomy} {\bf 72}, 83 (2019).



\end{thebibliography}
\end{document}